\documentclass{ws-fnl}
\usepackage{hyperref}
\usepackage{cite}
\usepackage{amsmath}

\begin{document}

\markboth{Alexander Vidybida and Olha Shchur}{Firing statistics of spiking neuron with delayed fast inhibitory feedback}

\catchline{}{}{}{}{}

\title{\bf RELATION BETWEEN FIRING STATISTICS OF SPIKING NEURON WITH DELAYED FAST INHIBITORY FEEDBACK AND WITHOUT FEEDBACK}

\author{\footnotesize ALEXANDER VIDYBIDA and OLHA SHCHUR}

\address{Bogolyubov Institute for Theoretical Physics, Metrologichna str., 14-B,\\
Kyiv, 03680,
Ukraine\\
vidybida@bitp.kiev.ua}

\maketitle

\begin{history}
\received{(received date)}
\revised{(revised date)}
\end{history}

\begin{abstract}
We consider a class of spiking neuronal models, defined by a set of conditions
typical for basic threshold-type models, such as the leaky integrate-and-fire
or the binding neuron model and also for some artificial neurons.
A neuron is fed with a Poisson process. Each output impulse is applied to the neuron itself after a 
finite delay $\Delta$. This impulse acts as being delivered through a fast {\em Cl}-type inhibitory synapse.
    We derive a general relation which allows calculating exactly the 
probability density function (pdf) $p(t)$ of output interspike
 intervals of a neuron with feedback based on known pdf $p^0(t)$
  for the same neuron without feedback and on the properties of the
  feedback line (the  $\Delta$ value).
  Similar relations between corresponding moments are derived. 

Furthermore, we prove that initial segment of pdf $p^0(t)$
for a neuron with a fixed threshold level is the same for any neuron
satisfying the imposed conditions and is completely determined by the input stream. For the 
Poisson input stream, we calculate that initial segment exactly and, based on it, 
obtain exactly the initial segment of pdf $p(t)$ for a neuron with feedback.
That is the initial segment of $p(t)$ is model-independent as well.
The obtained expressions are checked by means of Monte Carlo simulation. The course
of $p(t)$ has a pronounced peculiarity, which makes it impossible to approximate
$p(t)$ by Poisson or another simple stochastic process.

\end{abstract}

\bibliographystyle{ws-fnl.bst}

\section{Introduction}

In the theory of neural coding, the rate coding paradigm was dominating for a long time,
\cite{Adrian1928,Knight1972,Shadlen1994}. In the framework of this paradigm, the essential
neural signal is the mean number of impulses/spikes generated during some reference period,
but not their exact position in time. If the temporal position of spikes does not matter,
then it is natural to represent the seemingly random spike trains, produced by neurons, 
as Poisson processes, maybe with variable intensity, \cite{Shadlen1998,DOnofrio2016}. In the course
of gathering experimental data, it appeared that in some situations the temporal structure
of spike trains does have an essential role. E.g., this is valid for echolocation, \cite{Carr1990},
early vision, \cite{VanRullen2001}, motor control, \cite{Welsh1995}, late visual perception,
\cite{Amarasingham2006}. A further, more rigorous examination of experimental data revealed
that Poisson statistics, which does not have any temporal structure, 
is not suitable for modeling neuronal activity in some other brain areas, \cite{Maimon2009,Averbeck2009},
see also \cite{DiCrescenzo2005}.

What might be the reason for appearing a temporal structure in neuronal spike trains?
The two evident reasons are: 1) The threshold-type reaction to a stimulus; 2) A feedback
presence, either direct or indirect, through intermediate neurons. 
Due to 1), more than a single input impulse is required for triggering. This makes
improbable to get a short output ISI, while for Poisson distribution the shortest ISIs
are most probable. The second one may produce peculiarities in the pdf course 
for ISI length comparable with
the feedback delay time. This way, a fine temporal structure may appear in the spike trains
even if primary stimulation is due to Poisson process. One more reason for a temporal structure
in neuronal activity might be adaptation of any kind, e.g. 
\cite{Pirozzi2017,Carfora2016,Buonocore2014}.

In this paper, we study the latter possibility.
Namely, we consider a neuron with a fast inhibitory feedback and try to figure out what influence
the feedback presence may have on statistical properties of its activity.
Mathematically, we derive the ISI probability density function (pdf) for a neuron
with feedback from its ISI pdf without feedback and the feedback line properties.

In the previous paper \cite{Vidybida2015a} the required relation has been obtained
for the case when the feedback is excitatory and instantaneous, 
which allows calculating any one of the three pdfs, namely for stimulus, for 
output stream without feedback and for output stream with instantaneous feedback,
provided the other two are given.

In this paper,
we consider the case of fast inhibitory feedback with a \emph{non-zero} delay. Biological justification of
this case is given in Sec. \ref{Bjust}.
For this case, we obtain general relation allowing to calculate the ISI pdf $p(t)$
of a neuron with feedback based on known pdf $p^0(t)$ for the same neuron without feedback
stimulated with the same input Poisson stochastic process, see Eqs. (\ref{fscases}),
(\ref{ptfin}). The general relation is obtained for a class of neuronal models,
which includes the leaky integrate-and-fire and the binding
neuron  models\footnote{Definition of the binding neuron model can be found in \cite{Vidybida2014}. See also\\ https://en.wikipedia.org/wiki/Binding\_neuron.}, see Sec. \ref{class}, below.

Further, we analyze the  pdf $p^0(t)$ of a neuron without feedback and discover
that for any neuronal model there exists initial interval $]0;T[$ of ISI values at which $p^0(t)$
does not depend on the neuronal model chosen, being completely defined by
the input stream. Choosing Poisson stream as input, we calculate exactly
that model-independent initial segment of $p^0(t)$, see Sec. \ref{p0Poisson}.
This allows us to calculate exactly the initial segment of $p(t)$, 
which is model-independent as well, see Sec. \ref{th2}. A peculiarity in the $p(t)$
course for $t$ close to the delay time is clearly seen in Fig. \ref{LIF_graph}.

It appeared that the initial segment found for $p(t)$ is enough to express statistical moments of $p(t)$
through moments of $p^0(t)$, see  Sec. \ref{mome}, Eq. (\ref{moments}).
In particular, we obtain in our approach the model-independent relation between the mean ISI
of a neuron with and without feedback, which was known before for the binding
neuron model only, see Eq. (\ref{mean}).

Finally, we check the exact expressions found
for $p(t)$ by means of Monte Carlo simulation for 
the LIF neuron model, see Fig. \ref{LIF_graph}.

\section{Methods}
\subsection{Class of neuronal models}\label{class}

The neuronal state at the moment $t$ is described by the depolarization voltage
$V(t)$. If zero potential is chosen at the outer space of excitable membrane,
then at the resting state $V \sim -70$ mV. In order to simplify expressions, 
we consider biased by 70 mV values of $V$. In this case, 
$V=0$ at the resting state and depolarization is positive. 
This is similar (but not exactly the same) as in \cite{Hodgkin1952}.
 The input impulse increases the depolarization
voltage by $h$:
\begin{equation}\label{h}
V(t)\quad\to\quad V(t)+h,
\end{equation}
where $h>0$.

The input impulse decay is governed by a function $y(u)$,
which is different for different neuronal models. It means that
if the first (and single) impulse is received at the moment $t$, then for any $u>0$
\begin{equation}\label{decay}
V(t+u)=V(t)+hy(u).
\end{equation}

For instance, if we consider the LIF model than
\begin{equation}\label{fLIF}
y(u)=e^{-\frac{u}{\tau}},
\end{equation}
where $\tau$ is the relaxation time.

The neuron is characterized by a firing threshold value $V_0$: as soon as $V(t)>V_0$,
the neuron generates a spike and $V(t)$ becomes zero.

Instead of specifying any concrete neuronal model (through specifying $V_0$, $h$ and $y(u)$), we consider a class of neuronal models, which (without feedback) satisfy the following conditions:
\begin{itemize}
\item Cond0: Neuron is deterministic: Identical stimuli elicit identical 
spike trains from the same neuron.
\item Cond1: Neuron is stimulated with input Poisson stochastic process of excitatory impulses. 
\item Cond2: Just after firing, neuron appears in its resting state.
\item Cond3: The function $y(u)$, which governs decay of excitation, 
see Eq. (\ref{decay}), is continuous and satisfies the following conditions:
\begin{equation}\label{decrease}
\begin{split}
y(0)=1,\\
0<u_1<u_2\quad\Rightarrow\quad y(u_1)\geq y(u_2).
\end{split}
\end{equation}
The equality sign stays for the perfect integrator.
\item Cond 4: The pdf for output ISIs, $p^0(t)$, where $t$ means an ISI duration, exists 
together with all its moments.

\end{itemize}


Notice that we do not assume in the Eq. (\ref{h}) and Cond1, above,
 that the $h$ is the same for all input impulses.
Actually, the $h$ may depend on the number of input impulse,
or on the corresponding input ISI either deterministically, or stochastically,
as in renewal reward processes. 

We keep this freedom as regards the input stimulation up to the Sec. \ref{mome}, below,
where the general expressions are obtained.
For concrete cases, used as examples, we assume that the height of input impulse $h$
is the same for all input impulses. In this case,
the whole set of neuronal models, satisfying the above conditions
with different $V_0$ and $h$, can be decomposed into 
a set of
disjoint subsets numbered by $n=1,2,\dots$ by means of the following relation:
\begin{equation}\label{trn}
(n-1)h\le V_0<nh.
\end{equation}
If a model satisfies (\ref{trn}) with some $n$, then that neuron is said to have
threshold $n$. Indeed, in this case, $n$ specifies the minimal number of input impulses
necessary for triggering.\footnote{Actually, the threshold expressed in terms of membrane voltage is
always the same, but it is the height of input impulse $h$ which varies for the different subsets.}

\subsection{Type of feedback}
\begin{figure} [h]
\unitlength=0.9mm
\begin{center}
        \includegraphics[width=0.69\textwidth]{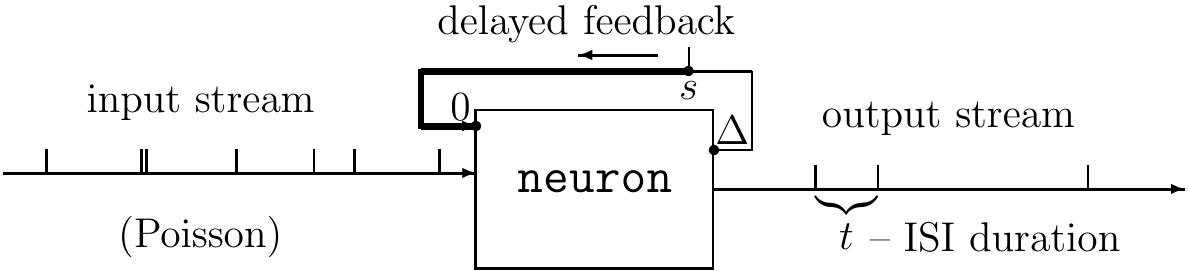}
\end{center}
\caption{\label{BNDwF} Neuron with delayed feedback. As {\large\tt neuron} in the figure
we consider any neuronal model, which satisfies the set of conditions Cond0 - Cond4, above.}
\end{figure}

The neuron is equipped with a delayed feedback line, see Fig. \ref{BNDwF}.
We expect that the feedback line has the following properties:
\begin{itemize}
\item Prop1: The time delay in the line is $\Delta>0$.
\item Prop2: The line is able to convey no more than one impulse.
\item Prop3: The impulse conveyed to the neuronal input is the fast {\em Cl-}type inhibitory
impulse. This means that after receiving such an impulse, the neuron appears 
in its resting state and the impulse is immediately forgotten.
\end{itemize}
\subsection{Biological justification}\label{Bjust}
It is known that neurons can form synaptic connections with their own body or dendrites.
Synapses of this type are called "autapses". For inhibitory neurons see \cite{Bekkers1998,Bacci2003,Bacci2004,Smith2002}.

There are two types of inhibitory currents initiated by two types of synapses.
Those currents are created by $Cl^-$ and $K^+$ ions.
The $K^+$ currents have rather slow kinetics, see \cite{Benardo1994,Bacci2004,Storm1988a,Storm1990}
  with the rise time ranging from tens to hundreds of
milliseconds and the decay constant between hundreds of milliseconds to minutes.
On the other hand, the $Cl^-$ current rise time is below 5 ms and the decay constant is up to 25  ms,
\cite{Benardo1994}. Having this in mind, and taking into account that the $Cl^-$ reversal potential
is equal to the rest potential, we model the $Cl^-$ action as immediately shunting any
excitation present in the neuron and doing nothing if the neuron is in its resting state.
This explains the Prop3, above. Hardware-based artificial neurons, like used in  \cite{Rossell2012,Wang2013},
can be covered by our approach as well, provided that Cond0-Cond4 and Prop1-Prop3,
above, are satisfied.

\section{Results}
\subsection{Pdf: general relation}
If the feedback line conveys  an impulse, let $s$ denote the time necessary
for that impulse to reach the end of the line and act upon the neuron.
Below, we call $s$ "time to live".
Notice that at the beginning of any ISI the line is never empty.
Let $f(s)$ denote the distribution of $s$ at the beginning of ISI in the stationary regime.
For calculating the pdf we introduce the conditional probability density $p(t|s)$,
which gives the probability to get the ISI $t$ units of time long provided that
at its beginning the line bears an impulse with the time to live $s$.
 In the stationary regime, required ISI pdf $p\left(t\right)$  can be calculated as follows:
\begin{equation}\label{pt}
p(t)=\int\limits_0^\Delta ds\: p(t|s) f(s).
\end{equation}

The function $p(t|s)$ for the fast {\em Cl-}type inhibitory feedback has been found
previously in \cite{Vidybida2017}. It looks as follows:
\begin{equation}\label{pts}
p(t|s)=\chi(s-t)p^0(t)+P^0(s)p^0(t-s),
\end{equation}
where
\begin{equation}\label{P0}
P^0(s)=1-\int\limits_0^s dt\: p^0(t),
\end{equation}
 $p^0(t)$ is the pdf without feedback and $\chi(t)$ is the Heaviside step function.

The pdf $f(s)$ satisfies the following equation, see \cite{Vidybida2015}:
\begin{equation}\label{fequat}
f\left(s\right)=\int\limits_s^\Delta ds'\: p^0\left(s'-s\right)f\left(s'\right)+
\delta\left(s-\Delta\right)\int\limits_0^\Delta ds'\: P^0\left(s'\right)f\left(s'\right).
\end{equation}

It appears from (\ref{pt},\ref{pts},\ref{fequat}) that it is only the exact expression 
for pdf without feedback $p^0(t)$ which is needed for finding the pdf with fast {\em Cl-}type 
inhibitory feedback.

It follows from (\ref{fequat}) that the $f\left(s\right)$ should have the following
form:
\begin{equation}\label{fs}
f\left(s\right)=g\left(s\right)+a\delta\left(s-\Delta\right),
\end{equation}
where $a > 0$ and $g(s)$ is bounded and vanishes out of interval $]0; \Delta]$.
After substituting Eq. (\ref{fs}) into (\ref{fequat}) and separating terms
with and without $\delta$-function, we obtain the following 
 system of integral equations for unknown $a$ and $g(s)$:
\begin{equation}\label{fscases}
\begin{cases}
a=\int\limits_0^\Delta ds'\: P^0(s')g(s')+aP^0(\Delta);\\
g(s)=\int\limits_s^\Delta ds'\: p^0(s'-s)g(s')+ ap^0(\Delta-s).
\end{cases}
\end{equation}

After substituting (\ref{pts},\ref{fs}) into (\ref{pt}) and taking into account  
that $p^0(t)$ equals to zero for negative argument, one can obtain the following:
\begin{equation}\label{ptfin}
p\left(t\right)=\begin{cases}
\int\limits_0^t ds\: P^0\left(s\right) p^0\left(t-s\right) g\left(s\right)
+p^0\left(t\right)\left(\int\limits_t^\Delta ds\: g\left(s\right)
+a\right),\;t<\Delta\\
aP^0\left(\Delta\right) p^0\left(t-\Delta\right)+
\int\limits_0^\Delta ds\: P^0\left(s\right) p^0\left(t-s\right) g\left(s\right),\;
t>\Delta.
\end{cases}
\end{equation}

This gives the straightforward algorithm for finding $p(t)$ from known $p^0(t)$:
\begin{enumerate}
\item[a:] Calculate $P^0(t)$ according to Eq. (\ref{P0}).
\item[b:]  Substitute $P^0(t)$ and
$p^0(t)$ into Eq. (\ref{fscases}) and find $a$ and $g(s)$.
\item[c:]  Substitute all 
into Eq. (\ref{ptfin}) and take the integrals.
\end{enumerate}

 Below (Sec. \ref{p0Poisson} -- \ref{th2}) we illustrate this in a situation when $p^0(t)$ is known exactly.

\subsection{Moments of pdf: general relations}\label{mome}
Denote $W_n$ and $W_n^0$ the $n$th moment of $p(t)$ and $p^0(t)$ respectively.
Using (\ref{ptfin}) within different
time domains $[0;\Delta[$ and $[\Delta;\infty[$, the moments of $p(t)$ can be found as follows:
\begin{equation}\nonumber
\begin{split}
W_n=\int\limits_0^{\infty}dt\: t^n p(t) \\ = \int\limits_0^\Delta dt\:t^n
\int\limits_0^t ds\:P^0(s) p^0(t-s) g(s)
+\int\limits_0^\Delta dt\:t^n p^0(t) \left(\int\limits_t^\Delta ds\:g(s) + a\right)\\
 + a P^0(\Delta) \int\limits_\Delta^{\infty} dt\:t^n p^0(t-\Delta)
+\int\limits_\Delta^{\infty} dt \:t^n \int\limits_0^\Delta  ds\: P^0(s) p^0(t-s) g (s)\\
=A_1+A_2+A_3+A_4.
\end{split}
\end{equation} 

Consider firstly $A_3$.
After replacing the
 integration variable $t$ by $(t-\Delta)$,
one can obtain the following:
\begin{equation}\nonumber
A_3=aP^0(\Delta)\int\limits_0^{\infty}dt \:(t+\Delta)^n p^0(t).
\end{equation} 

Using the moments' definition for pdf without feedback, $W_n^0$, and
taking into account that $W^0_0$ is a normalization coefficient and equals $1$, one has:
\begin{equation}\nonumber
A_3=aP^0(\Delta) \sum_{k=0}^n \binom{n}{k} W_k^0 \Delta^{n-k}.
\end{equation} 

In $A_4$ we change the integration order:
\begin{equation}\nonumber
A_4=\int\limits_0^\Delta ds\: P^0\left(s\right) g(s)
\int\limits_{\Delta}^{\infty}dt\:t^n p^0(t-s),
\end{equation} 
and then we replace the integration variable $t$ by $(t-s)$:
\begin{equation}\nonumber
A_4 =\int\limits_0^\Delta ds\: P^0(s) g (s)
\int\limits_{\Delta-s}^{\infty} dt\:(t+s)^n p^0(t).
\end{equation} 

Notice that $\int\limits_{\Delta-s}^{\infty}dt=\int\limits_0^{\infty}dt-
\int\limits_0^{\Delta-s}dt$. Using this, normalization of $p^0(t)$ and definitions
for $W_n^0$, one gets:
\begin{equation}\label{a4}
A_4=\int\limits_0^\Delta ds\: P^0(s)g(s) \sum_{k=0}^n \binom{n}{k} W_k^0 s^{n-k}
-\int\limits_0^\Delta ds\: P^0(s) g (s) \int\limits_0^{\Delta-s}dt \:(t+s)^n p^0(t).
\end{equation} 

Now,  change the integration order in $A_1$. This gives the same expression
as the second term in (\ref{a4}). Finally, after cancellation we have:
\begin{equation}\label{moments}
\begin{split}
W_n=
\int\limits_0^\Delta dt\:t^n p^0(t) \left(\int\limits_t^\Delta ds\:g(s) + a\right)\\
+ \sum_{k=0}^n \binom{n}{k} W_k^0 \left(a P^0(\Delta) \Delta^{n-k}
+\int\limits_0^\Delta ds\: g(s)P^0(s)s^{n-k} \right).
\end{split}
\end{equation} 

\subsection{Initial segment of $p^0(t)$ for Poisson input}\label{p0Poisson}

Here we consider a situation with the feedback line removed.
Obviously, according to (\ref{decay}), (\ref{decrease}), the neuronal firing
may happen only at the moment of receiving an input impulse. 
Expect that Eq. (\ref{trn}) is valid with some fixed $n>1$ for our neuronal model.
Let the last firing happen at the moment 0.
After the last firing, the neuron requires at least $n$
impulses in order to be triggered again. Denote as $t$ the interval between the 
last and the next firing. The probability density function of this $t$
is just\footnote{Notice that neuron without feedback has a renewal stochastic process
as its output.}
 $p^0(t)$. The first firing at the moment $t$ may be triggered
by the input impulse \#$n$, or greater. But for any neuron satisfying Cond0 - Cond4
there exists an initial interval of $t$ values, $[0;T_n]$. Those values can be achieved 
by triggering {\em only} due to the input impulse \#$n$. The latter is clear for very
small values of $t$. For infinitesimally small $t$, the input impulse \#$n$
ensures excitation since $nh>V_0$, living no chances for the impulse \#$(n+1)$ to trigger
the first firing. $T_n$ is the maximum of such $t$ 
(ISI, which is achievable by the $n$th input impulse only).
The situation here is similar to discussed earlier for the 
LIF neuron in \cite[Theorem 2]{Vidybida2016}.

To figure out the exact length of that initial interval,
consider the situation when the first $(n-1)$ input impulses are obtained
immediately after the last firing. Now, $T_n$ is the maximum value of $t$, such
that impulse \#$n$ delivered at $t$ still triggers.

At the moment $T_n$, the first $(n-1)$ impulses ensure excitation $(n-1)hy(T_n)$.
Now,  $T_n$ can be found from the following equation:
\begin{equation*}
(n-1)hy(T_n) =V_0-h,
\end{equation*}
or
\begin{equation*}
T_n=y^{-1}\left(\frac{V_0-h}{(n-1)h}\right),
\end{equation*}
where the function $y^{-1}$ is the inverse of $y$. For the perfect integrator $T_n=\infty$.

According to the definition of $T_n$, any $n$ input impulses received within
$]0;T_n]$ evoke firing at the moment of receiving the last one.
Therefore, the pdf $p^0(t)\,dt$ for $t\in [0;T_n]$ is the probability to receive 
$(n-1)$ input impulses within
$]0;t[$ and the impulse \#$n$ within $[t;t+dt[$, whatever the $y(u)$ might be.
In particular, for Poisson input with intensity $\lambda$, we have
 on the initial segment a $\gamma$-distribution: 
\begin{equation}\label{p0Pu}
p^0(t)\,dt=e^{-\lambda t}\frac{(\lambda t)^{n-1}}{(n-1)!}	\lambda dt,\quad t\le T_n.
\end{equation}

This is in concordance with \cite[Eq. (21)]{Vidybida2016}, \cite[Eq. (3)]{Vidybida2007}, 
where $p^0(t)$ within $]0;T_2]$ is found for the LIF and the binding neuron, respectively.

It is worth noticing that features of a concrete neuronal model are
present in Eq. (\ref{p0Pu}) only through the value of $T_n$  The time course of
$p^0(t)$ within $]0;T_n]$ does not depend on neuron's physical properties (the
 manner of decaying of input stimuli governed by function $y(u)$ in (\ref{decay})), 
 but is completely determined by the input stochastic process. 
 
As regards the $T_n$ value, 
for the LIF neuron model $y(u)$ is defined by the Eq. (\ref{fLIF}), from which
we have
\begin{equation}\label{TnLIF}\nonumber
T_n=\tau\log\left(\frac{(n-1)h}{V_0-h}\right).
\end{equation}

\subsection{Distribution of times to live}

Here and further we expect that 
\begin{equation}\label{condDT}
\Delta < T_n.
\end{equation}
This allows us to obtain exact expressions in the case of Poisson stimulation 
without specifying an exact neuronal model.

In order to find the distribution $f(s)$ for time to live, one has to 
solve the system (\ref{fscases}). It is clear from (\ref{fscases}) that it
determines the pair $(a, g(s))$ only up to arbitrary coefficient.
In order to fix that uncertainty, we use the normalization condition:
\begin{equation}\label{norm}
1=\int\limits_0^\Delta ds\: g\left(s\right)+a.
\end{equation}
 Let us rewrite the second equation in (\ref{fscases})
by replacing the unknown function with $\tilde g(s)={g(s)}/{a}$:
\begin{equation}\label{gequat}
\tilde g(s)=\int\limits_s^\Delta ds'\: p^0(s'-s)\tilde g(s')+
p^0(\Delta-s).
\end{equation}

Due to (\ref{condDT}), we may use Eq. (\ref{p0Pu}) with certain $n$ in Eq. (\ref{gequat}) 
in order to find the bounded part of the distribution of times to live $f(s)$. 
For a neuron with threshold $n=2$ we have:
\begin{equation}\nonumber
\tilde g\left(s\right)=\lambda^2 e^{\lambda s}\int\limits_s^\Delta ds'\: e^{-\lambda s'} s'
\tilde g\left(s'\right)-\lambda^2 e^{\lambda s}s\int\limits_s^\Delta ds'\: e^{-\lambda s'}
\tilde g\left(s'\right)+\lambda^2 e^{-\lambda\left(\Delta-s\right)}\left(\Delta-s\right).
\end{equation}

The solution is:
\begin{equation}\label{tildeg}
\tilde{g}\left(s\right)=\dfrac{\lambda}{2}\left(1- e^{-2\lambda\left(\Delta-s\right)}\right).
\end{equation}

From the Eq. (\ref{norm})
\begin{equation}\label{a}
a=\dfrac{4e^{2\lambda\Delta}}{1+e^{2\lambda\Delta}\left(2\lambda\Delta+3\right)}.
\end{equation}

Interesting that the last two equations are the same as in \cite[Eq. (15-16)]{Vidybida2008a},
 where those are obtained by another method for the binding neuron model with threshold 2.

\subsection{Initial segment of $p(t)$ for threshold 2}\label{th2}
We can find the pdf $p(t)$ for threshold 2 within
$]0;\Delta]$ after substituting (\ref{p0Pu}) with $n=2$, (\ref{tildeg}) and (\ref{a}) 
into the first equation in (\ref{ptfin}):
\begin{equation}\label{p1}
\begin{split}
p\left(t\right)=\dfrac{2\lambda e^{-\lambda t}}{3+2\Delta\lambda+e^{-2\lambda\Delta}}\\
\left(\dfrac{1}{6}\lambda^3 t^3-\dfrac{1}{2}\lambda^2 t^2+\lambda^2t\Delta +
\lambda t\left(\dfrac{3}{2}+\dfrac{1}{4}e^{-2\lambda\Delta}+
\dfrac{1}{4}e^{-2\lambda\left(\Delta-t\right)}\right)\right), \quad t<\Delta.
\end{split}
\end{equation}

This is in concordance with what was obtained in \cite[Eq. (21)]{Vidybida2013}
for the binding neuron. Here we have proven that this same distribution of ISIs
is valid for all neuronal models satisfying the conditions of Sec. \ref{class}, above.

Within $]\Delta;T_2[$, $p(t)$ is obtained by integrating the second equation from
(\ref{ptfin}):
\begin{equation}\label{p2}
\begin{split}
p\left(t\right)=\dfrac{2\lambda e^{-\lambda t}}{3+2\Delta\lambda+e^{-2\lambda\Delta}}\\
\left(\lambda t\left(\dfrac{1}{2}\lambda^2\Delta^2+\dfrac{5}{2}\lambda\Delta+
\dfrac{7}{4}+\dfrac{1}{4}e^{-2\lambda\Delta}\right)-\dfrac{1}{3}\lambda^3\Delta^3
-2\lambda^2\Delta^2-2\lambda\Delta\right), \, \Delta<t<T_2.
\end{split}
\end{equation}

Since $T_2=\infty$ for the perfect integrator, in that case, we have obtained the desired
pdf within entire positive time semi-axis.

\begin{figure}
  \unitlength=1mm
    \includegraphics[width=0.47\textwidth]{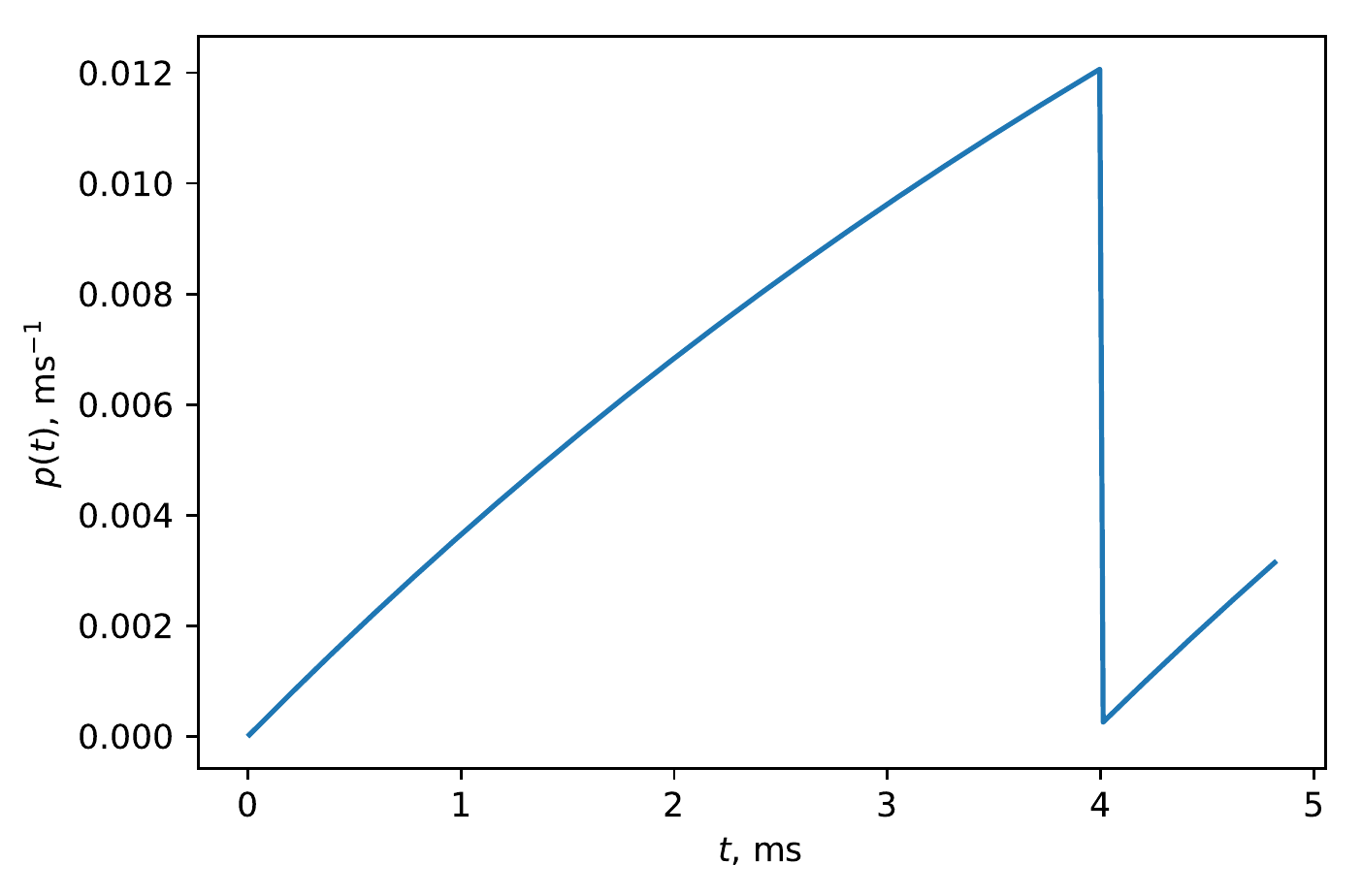}
    \hfill
    \includegraphics[width=0.47\textwidth]{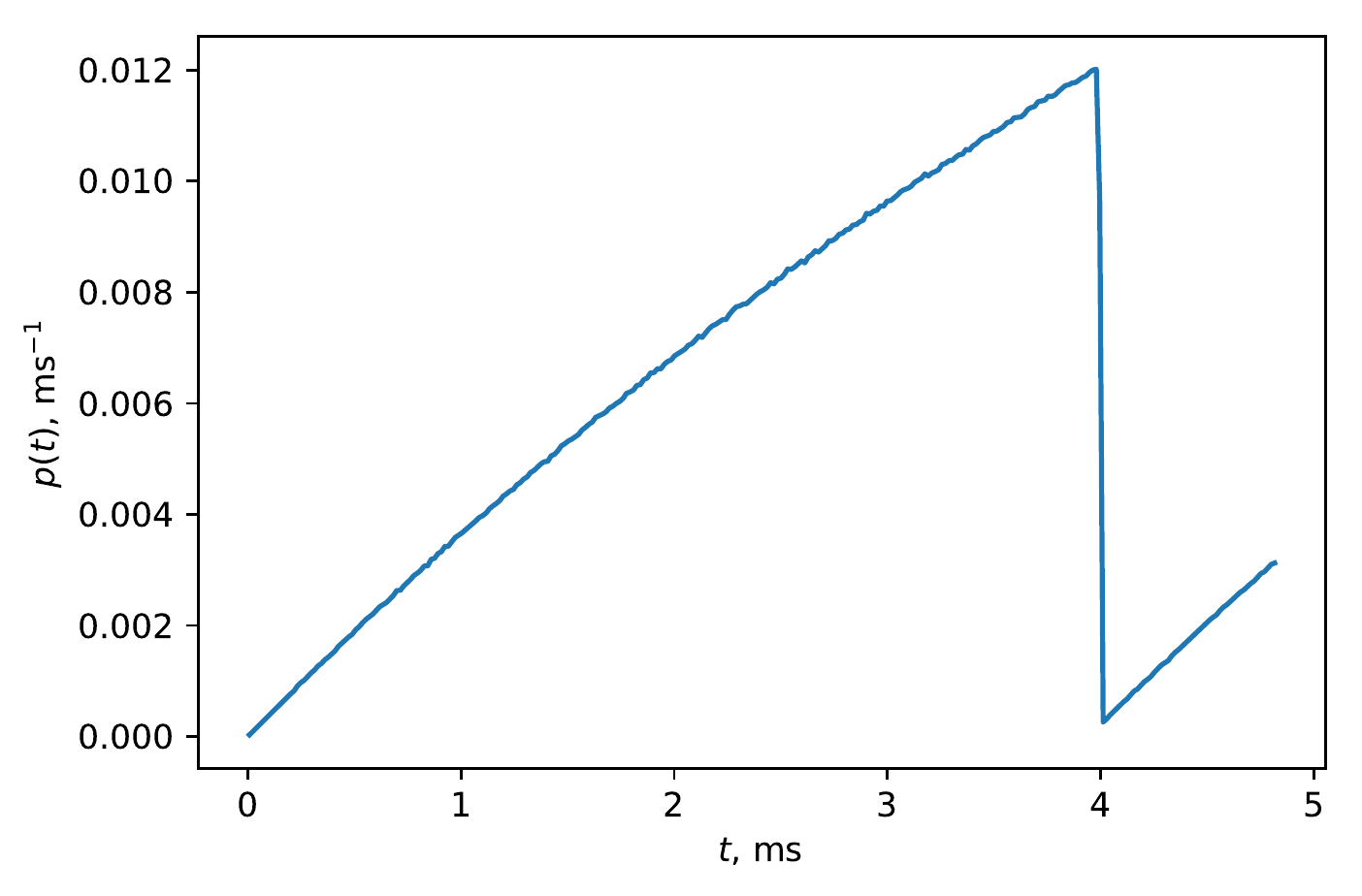}
  \hfill 
  \caption{\label{LIF_graph} Example of ISI PDF for the LIF neuron with threshold 2. 
  Left --- calculation in accordance with Eqs. (\ref{p1}), (\ref{p2}).
Right --- Monte Carlo simulation (used 1000000000 output ISIs). For both panels:
$\tau=20$ ms, $V_0=20$ mV,  $h=11.2$ mV, 
($\tau$, $V_0$ and $h$ are defined in Sec. \ref{class})
$\Delta=4$ ms, $\lambda=62.5$ s$^{-1}$.}
\end{figure}

\subsection{The first and the second moments for threshold 2}
To find the moments of pdf $p(t)$ one should substitute (\ref{p0Pu}) with $n=2$,
(\ref{tildeg}) and (\ref{a})
into (\ref{moments}). Notice that in (\ref{moments}), one needs to know $p^0(t)$
only for $t\le\Delta$, which for Poisson input and restriction (\ref{condDT})
 is given in (\ref{p1}).

Thus, after taking all the integrals, we obtain the following
expression for mean $W_1$ of $p(t)$:
\begin{equation}\label{mean}
W_1=a(W_1^0+\Delta)=
\dfrac{4e^{2\lambda\Delta}}{1+e^{2\lambda\Delta}\left(2\lambda\Delta+3\right)}
\left(W_1^0+\Delta\right).
\end{equation} 
This is in concordance with \cite[Eq. (23)]{Vidybida2013} obtained for the binding
neuron with threshold 2. The output intensity can be now found as
$$
\lambda^{out}=\frac{1}{W_1}.
$$

Similarly, one can obtain the following for the second moment $W_2$:
\begin{equation}\label{W2}
W_2=
\dfrac{2 \left(-1+2W_1^0 \lambda +8e^{\Delta  \lambda } (1-W_1^0\lambda )
+e^{2 \Delta  \lambda } \left(-7+6\lambda(W_1^0 
+\Delta) +2 W_2^0 \lambda ^2\right)\right)}
{\lambda^2\left(1+e^{2\lambda\Delta}\left(2\lambda\Delta+3\right)\right)}.
\end{equation} 

Thus to find the moments of pdf for the neuron with feedback it is enough to know the
corresponding moments for the neuron without feedback. For example, one can find
the output intensity for the LIF neuron with inhibitory feedback by means of substituting
the expression for mean without feedback,
\cite[Eq. (46)]{Vidybida2016}, into (\ref{mean}).


\section{Conclusions and discussion}

In this paper, we have derived general mathematical expressions, see Eq. (\ref{ptfin}), 
for calculating pdf  $p(t)$ of output ISI distribution for a neuron with delayed fast inhibitory feedback
stimulated with a Poisson stream of input impulses based on the pdf for
that same neuron without feedback. The expression found is valid for a class of neuronal models
defined by a set of natural conditions, see Cond0-Cond4 in Sec. \ref{class}. Standard threshold-type
models, like leaky integrate-and-fire model or binding neuron model, satisfy the  
conditions mentioned above. Similar general expressions are derived for moments of the pdf found, Sec. \ref{mome}.
In the case of Poisson input stimulus, we obtain a model-independent exact expression for the initial
segment of $p(t)$ and its moments, provided that a model is characterized by threshold 2, Eqs. 
(\ref{p1}) - (\ref{W2}).

The course of the pdf found, see Fig. \ref{LIF_graph}, has clearly seen peculiarity --- a jump 
for ISI $t=\Delta$, which excludes a possibility to describe the output ISI stream by a 
Poisson-like, or other simple distribution.  Our findings add to the discussion about such a possibility, see 
\cite{Maimon2009,Averbeck2009}.

\section*{Acknowledgements}
This research was supported by theme grant of department of physics and astronomy of NAS Ukraine "Dynamic
formation of spatial uniform structures in many-body system" PK
0118U003535.



\begin{thebibliography}{10}
\newcommand{\enquote}[1]{``#1''}

\bibitem{Adrian1928}
E.~D. Adrian, \emph{The Basis of Sensation, the Action of the Sense Organs}
  (Christophers, 1928).

\bibitem{Knight1972}
B.~W. Knight, \enquote{Dynamics of encoding in a population of neurons},
  \emph{The.Journal of General.Physiology.} \textbf{59} (1972) 734--766.

\bibitem{Shadlen1994}
M.~N. Shadlen and W.~T. Newsome, \enquote{Noise, neural codes and cortical
  organization}, \emph{Current Opinion in Neurobiology} \textbf{4} (1994)
  569--579.

\bibitem{Shadlen1998}
M.~N. Shadlen and W.~T. Newsome, \enquote{The variable discharge of cortical
  neurons: Implications for connectivity, computation, and information coding},
  \emph{The Journal of Neuroscience} \textbf{18} (1998) 3870--3896.

\bibitem{DOnofrio2016}
G. D'Onofrio and E. Pirozzi,
\enquote{Successive spike times predicted by a stochastic neuronal model with a 
variable input signal},
\emph{Mathematical Biosciences and Engineering} \textbf{13} (2016) 495--507.


\bibitem{Carr1990}
C.~E. Carr and M.~Konishi, \enquote{A circuit for deteection of interaural time
  differences in the brain stem of the barn owl}, \emph{The Journal of
  Neuroscience} \textbf{10} (1990) 3227--3246.

\bibitem{VanRullen2001}
R.~VanRullen and S.~Thorpe, \enquote{Rate coding versus temporal order coding:
  What the retinal ganglion cells tell the visual cortex}, \emph{Neural
  Computation} \textbf{13} (2001) 1255--1283.

\bibitem{Welsh1995}
J.~P. Welsh, E.~J. Lang, I.~Sugihara and R.~Llinás, \enquote{Dynamic
  organization of motor control within the olivocerebellar system},
  \emph{Nature} \textbf{374} (1995) 453--457.

\bibitem{Amarasingham2006}
A.~Amarasingham, T.~L. Chen, S.~Geman, M.~T. Harrison and D.~L. Sheinberg,
  \enquote{Spike count reliability and the poisson hypothesis}, \emph{The
  Journal of Neuroscience} \textbf{26} (2006) 801--809.

\bibitem{Maimon2009}
G.~Maimon and J.~A. Assad, \enquote{Beyond poisson: Increased spike-time
  regularity across primate parietal cortex}, \emph{Neuron} \textbf{62} (2009)
  426--440.

\bibitem{Averbeck2009}
B.~B. Averbeck, \enquote{Poisson or not Poisson: Differences in spike train
  statistics between parietal cortical areas}, \emph{Neuron} \textbf{62} (2009)
  310--311.

\bibitem{DiCrescenzo2005}
A. Di Crescenzo, B. Martinucci and E. Pirozzi,
\enquote{On the dynamics of a pair of coupled neurons subject to alternating input rates},
\emph{BioSystems} \textbf{79(1-3 SPEC. ISS.)} (2005) 109--116.


\bibitem{Buonocore2014}
A. Buonocore, L. Caputo, E. Pirozzi and M.~F. Carfora,
\enquote{Gauss-diffusion processes for modeling the dynamics of a couple of interacting neurons}
\emph{Math Biosci Eng} \textbf{11} (2014) 189--201.


\bibitem{Carfora2016}
M.~F. Carfora, E. Pirozzi, L. Caputo and A. Buonocore,
\enquote{A leaky integrate-and-fire model with adaptation for the generation of a spike train}
\emph{Mathematical Biosciences and Engineering} \textbf{13} (2016) 483--493.


\bibitem{Pirozzi2017}
Pirozzi,E.
\enquote{Colored noise and a stochastic fractional model for 
correlated inputs and adaptation in neuronal firing}
\emph{Biological Cybernetics}  (2017). https://doi.org/10.1007/s00422-017-0731-0



\bibitem{Vidybida2015a}
A.~K. Vidybida, \enquote{Relation between firing statistics of spiking neuron
  with instantaneous feedback and without feedback}, \emph{Fluctuation and
  Noise Letters} \textbf{14} (2015) 1550034.

\bibitem{Vidybida2014}
A.~K. Vidybida, \enquote{Binding neuron}, in \emph{Encyclopedia of information
  science and technology}, ed. M.~Khosrow-Pour (IGI Global, 2014), pp.
  1123--1134.

\bibitem{Hodgkin1952}
A.~L. Hodgkin and A.~F. Huxley, \enquote{A quantitative description of membrane current and its application to conduction and excitation in nerve}
\emph{Journal of Physiology} \textbf{117} (1952) 500--544.


\bibitem{Bekkers1998}
J.~M. Bekkers, \enquote{Neurophysiology: Are autapses prodigal synapses?},
  \emph{Current Biology} \textbf{8} (1998) R52--R55.

\bibitem{Bacci2003}
A.~Bacci, J.~R. Huguenard and D.~A. Prince, \enquote{Functional autaptic
  neurotransmission in fast-spiking interneurons: A novel form of feedback
  inhibition in the neocortex}, \emph{The Journal of Neuroscience} \textbf{23}
  (2003) 859--866.

\bibitem{Bacci2004}
A.~Bacci, J.~R. Huguenard and D.~A. Prince, \enquote{Long-lasting
  self-inhibition of neocortical interneurons mediated by endocannabinoids},
  \emph{Nature} \textbf{431} (2004) 312--316.

\bibitem{Smith2002}
T.~C. Smith and C.~E. Jahr, \enquote{Self-inhibition of olfactory bulb
  neurons}, \emph{Nature Neuroscience} \textbf{5} (2002) 760--766.

\bibitem{Storm1988a}
J.~F. Storm, \enquote{Four voltage-dependent potassium currents in adult 
hippocampal pyramidal cells},
\emph{Biophys. J.} \textbf{53} (1988) 148a.

\bibitem{Storm1990}
J. Storm, \enquote{Potassium currents in hippocampal pyramidal cells}
\emph{Progress in Brain Research} \textbf{83} (1990) 161--187.

\bibitem{Benardo1994}
L.~S. Benardo, \enquote{Separate activation of fast and slow inhibitory
  postsynaptic potentials in rat neocortex in vitro}, \emph{Journal of
  Physiology} \textbf{476.2} (1994) 203--215.

\bibitem{Rossell2012}
J.~L. Rosselló, V.~Canals, A.~Morro and A.~Oliver, \enquote{Hardware
  implementation of stochastic spiking neural networks}, \emph{International
  Journal of Neural Systems} \textbf{22} (2012) 1250014.

\bibitem{Wang2013}
R.~Wang, G.~Cohen, K.~M. Stiefel, T.~J. Hamilton, J.~Tapson and A.~van Schaik,
  \enquote{An FPGA implementation of a polychronous spiking neural network with
  delay adaptation}, \emph{Frontiers in Neuroscience} \textbf{7}.

\bibitem{Vidybida2017}
A.~K. Vidybida, \enquote{Fast {\em Cl}-type inhibitory neuron with delayed feedback
  has non-markov output statistics}, \emph{Biological.Cybernetics.} .

\bibitem{Vidybida2015}
A.~K. Vidybida, \enquote{Activity of excitatory neuron with delayed feedback
  stimulated with poisson stream is non-markov}, \emph{Journal of Statistical
  Physics} \textbf{160} (2015) 1507--1518.

\bibitem{Vidybida2016}
A.~K. Vidybida, \enquote{Output stream of leaky integrate-and-fire neuron
  without diffusion approximation}, \emph{Journal of Statistical Physics}
  \textbf{166} (2016) 267--281.

\bibitem{Vidybida2007}
O.~Vidybida, \enquote{Output stream of a binding neuron}, \emph{Ukrainian
  Mathematical Journal} \textbf{59} (2007) 1819--1839.

\bibitem{Vidybida2008a}
A.~K. Vidybida, \enquote{Output stream of binding neuron with delayed
  feedback}, in \emph{14th International Congress of Cybernetics and Systems of
  WOSC, Wroclaw, Poland, September 9-12, 2008}, eds. J.~Józefczyk, W.~Thomas
  and M.~Turowska (Oficyna Wydawnicza Politechniki Wroclawskiej, 2008), pp.
  292--302.

\bibitem{Vidybida2013}
A.~K. Vidybida and K.~G. Kravchuk, \enquote{Firing statistics of inhibitory
  neuron with delayed feedback. i. output isi probability density},
  \emph{BioSystems} \textbf{112} (2013) 224--232.

\end{thebibliography}

\end{document}